\documentclass[12pt]{iopart}

\usepackage{graphics,epsfig}
 
\newcommand{\be}{\begin{eqnarray}}
\newcommand{\ee}{\end{eqnarray}}
 
\begin{document}


\title{Is the Stillinger and Weber decomposition relevant for coarsening 
models?}

\author{A. Crisanti$^{1}$, F. Ritort$^{2}$, A. Rocco$^{1,3}$, 
and M. Sellitto$^{4}$} 

\address{$^{1}$ Dipartimento di Fisica, Universit\`a di
Roma ``La Sapienza'', P.le Aldo Moro 2, I-00185 Roma, Italy and 
Istituto Nazionale Fisica della Materia, Unit\`a di Roma.}

\address{$^{2}$ Departament FFN, Facultat de F\'{\i}sica, 
Universitat de Barcelona Avda. Diagonal 647, 08028 Barcelona, Spain.}

\address{$^{3}$ CWI, Postbus 94079, 1090 GB Amsterdam, The Netherlands.}

\address{$^{4}$ Abdus Salam International Centre for Theoretical Physics,
34100 Trieste, Italy.}



\begin{abstract}
We study three kinetic models with constraint, namely the
Symmetrically Constrained Ising Chain, the Asymmetrically Constrained
Ising Chain, and the Backgammon Model. All these models show
glassy behaviour and coarsening. We apply to them the 
Stillinger and Weber decomposition, and
find that they share the same configurational
entropy, despite of their different nonequilibrium dynamics.
We conclude therefore that the Stillinger and Weber decomposition 
is not relevant for this type of models.
\end{abstract}

\section{Introduction}

The description of the glassy dynamics remains an intriguing issue,
even after years of research \cite{review}. In this context considerable 
progress has been achieved by the introduction of {\em constrained kinetic 
Ising models}. 
In these models the {\em slowing down} of the dynamics is realized through 
the introduction of microscopic kinetic constraints, which serve
the purpose of preventing certain spins from being flipped. The
first proposal was made by Fredrickson and Andersen in 1984 \cite{fred}
in the attempt to provide a simple microscopic mechanism for 
understanding the purely dynamical transition predicted by the 
mode coupling theory. Along the same lines J\"ackle and Eisinger
\cite{jackle} later on modified that model, inserting a stronger constraint
which results in an exponential inverse temperature squared 
dependence for the relaxation time \cite{sollich}. More recently,
constrained Ising chains have also been considered as simple models for  
granular compaction \cite{dean}. As a matter of
fact, all these models show {\em glassy behaviour} in the sense that their 
relaxation times diverge when temperature is lowered \cite{crrs}. 
Their relaxation toward equilibrium proceeds through the coalescence of
domains of either up or down spins. This process is characterized by a
{\em growing length scale} (the average domain length), which drives the
system toward equilibrium and signals the {\em coarsening behaviour} of these 
models.

As it is well known, the description of the slow dynamics of either spin 
glasses or structural glasses rests on the idea of the exploration of the 
configuration space through thermal jumps ({\em activated dynamics}). The 
more the temperature is lowered, the more the system gets confined in 
localized regions of the phase space, pretty much as a golf ball gets trapped 
in the valleys of the green. Whether or not coarsening can be considered as 
another prototype process relevant to the description of the glassy dynamics 
is an open question. Probably a reasonable answer calls for the superposition 
of both processes, namely coarsening and activation. 

In the case of systems exhibiting mainly activated dynamics, an
interesting approach was proposed by Stillinger and Weber (SW) in the early
Eighties \cite{sw}. Their approach was based on the decomposition of the
configuration space into {\em valleys} 
on the basis of the topology of the potential energy landscape. To
each valley a label, called Inherent Structure (IS), is attached, and
the offequilibrium dynamics of the system is reduced to a dynamics
defined on the IS configurations. This projection technique has been
proven to be relevant to the glass transition in several cases, in
the domain of both potential \cite{skt,crisanti} and free energy 
\cite{marinari}. 

An interesting question, however, is whether or not this approach can be
straightforwardly applied to coarsening systems too. The question is
far from trivial because for these systems, on top of the
activated dynamics, there is also a geometrical constraint leading the
system to explore the configuration space along a coarsening path. 
In contrast to what happens for
purely activated dynamics, now, once the systems sits in some valley,
the choice of the next one to reach via a thermally activated jump
is not simply related to the number and dimensions of the neighboring
valleys, but is also driven by the constraint that the average domain
length must grow with time. Therefore the suspicion that the SW
decomposition may be not able to reproduce the offequilibrium behaviour
of these systems is legitimate and requires specific
attention. As we shall see, it seems that indeed the SW decomposition
does not capture the specific dynamics of coarsening models.

The paper is organized as follows. In the next section we shall
present the models under study, namely the Symmetrically and the 
Asymmetrically Constrained Ising Chain. The Backgammon Model will also be 
introduced for comparison. Then, in section III we shall analyse the 
response of these models to perturbations, discussing their 
fluctuation dissipation relations. In section IV we shall present the      
Stillinger and Weber decomposition and show that the corresponding 
configurational entropy does not account for the different dynamics 
of these systems. Finally in section V we shall draw some conclusions.

\section{The models} 
 
\subsection{The Symmetrically Constrained Ising Chain (SCIC)} 
 
The Symmetrically Constrained Ising Chain was first introduced by  
Fredrickson and Andersen \cite{fred} in 1984. The model is defined as follows: 
\numparts
\begin{eqnarray} 
E = - \sum_{i=1}^N \sigma_i \\ 
{\cal W} (\sigma_i \rightarrow 1 - \sigma_i) = \frac{1}{2} [2 - 
\sigma_{i-1} - \sigma_{i+1}] \; \min\{1,e^{-\beta \Delta E}\}
\label{SCIC} 
\end{eqnarray}
\endnumparts 
Here the variables $\sigma$'s are Ising-like spin variables, which can 
take up the values 0 (down spin) or 1 (up spin).    
The ordinary Glauber rule is defined on a 
restricted class of mobile spins, making thereby the dynamics of the 
model far from being trivial. More specifically, the constraint present in 
the transition probability makes the ordinary update possible only for 
those spins whose left or right first neighbour is found in the down  
state. For all the other spins the corresponding transition rate is 
zero. As a result, even though the energy of the system simply 
corresponds to a paramagnet in a field, its dynamics turns out to be 
much richer, and particularly the approach toward equilibrium is 
expected to show slow motion properties.     
 
\begin{figure}[h] 
\hbox{ 
\centerline{\epsfig{file=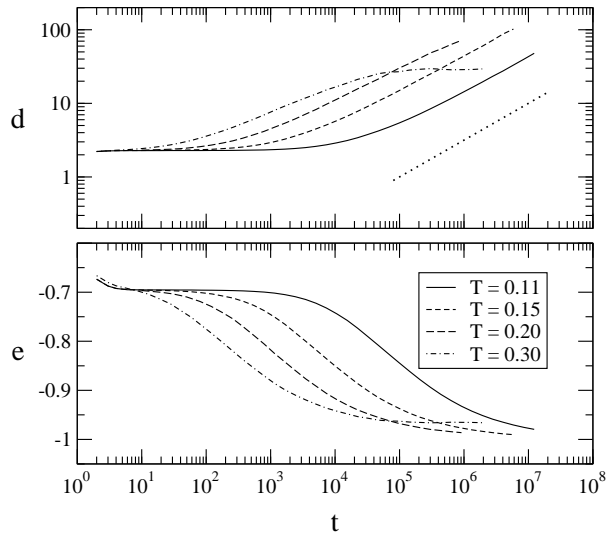,width=8cm}} 
} 
\caption{Relaxation of average domain length and energy in the SCIC 
at different temperatures. The dotted line corresponds to the power 
law $t^{1/2}$.}  
\label{FAde} 
\end{figure}

To get more insights into the relaxation properties of the model, let 
us discuss briefly its microscopic dynamics, as defined by (1).  
Starting from an initial condition where each spin is assigned 
randomly the value 0 or 1, there will be an initial situation, 
characterized by a time scale which will be specified in the following,  
where a quite fast growth of small {\em domains} of spins 
in the state up will occur. These domains will be separated by spins in 
the down state, which from now on will be called {\em defects}. After this 
initial phase, the equilibration of the system will proceed through the 
process of eliminating defects. This is where the constraint enters 
strongly into play. By definition of a defect, both its neighbors 
are in the up state, and therefore its flipping is forbidden. The 
only possibility of eliminating it will be to carry another defect 
({\em auxiliary defect}) to its right or left, forcing it to travel along 
one of the two adjacent domains. This process is clearly slow  
because the traveling of the auxiliary defect toward the original one 
will involve the overturning of up spins into down spins,  
with an increase of the energy of the system as determined by the 
Metropolis factor. Then it will become possible to flip the original 
defect, and, when flipped, one of the two adjacent domains will   
increase its length by one unit. 
To complete the process, we still need to make the auxiliary defect  
travel back to its original position.  
Once this situation is be achieved, the two 
original domains will have coalesced into a single one, with no 
other change in the chain of spins, and the energy will have indeed 
decreased. This is what we mean by {\em coarsening}. Note that this  
process will be slower and slower the closer the system is to  
equilibrium, since the domains of up spins get longer and longer  
with time. This is the origin of the glassy behaviour of the model.  
The relaxation of both average domain length and energy is shown in  
Fig.\ \ref{FAde} for different temperatures.  It is easy to show that 
regardless the dynamical rules the two quantities are related via 
$d = -e/(1+e)$ \cite{nota}, so the use of $d$ or $e$ is just a matter of taste.
 
The dynamics of the model is characterized by the    
existence of three relevant time scales.  
The first time scale has an Arrhenius behaviour and is strongly 
dependent on the constrained dynamics. It corresponds to the  
microscopic relaxation of one spin with its nearest neighbours (local 
equilibration) and can be evaluated to be $\tau_1 \sim \exp(\beta)$ 
\cite{follana}. Only for times larger than $\tau_1$ will nonequilibrium 
behaviour appear, with nonexponential relaxation and aging 
effects. Another relevant time scale is the equilibration time 
which can be estimated as $\tau_{\rm eq} \sim \exp(\lambda \beta)$ with 
$\lambda$ in the range $3 \ldots 4$. In fact a simple scaling analysis
shows that
$\lambda \sim 3$ \cite{crrs}. Finally one can also 
define a correlation time $\tau_{\rm corr} \sim \exp(2 \beta)$ as the integral 
of the equilibrium connected correlation function \cite{crrs}. This 
dependence on temperature is in agreement with the previous results of  
\cite{reiter,schulz}. Note in Fig.\ \ref{FAde} the initial plateau 
related to the time scale $\tau_1$ and the diffusive growth of the 
average domain length.  

\subsection{The Asymmetrically Constrained Ising Chain (ACIC)} 
 
\begin{figure}[h]  
\hbox{ 
\centerline{\epsfig{file=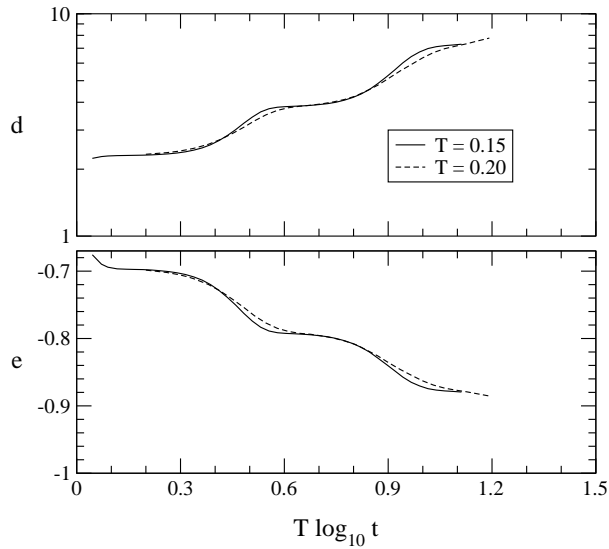,width=8cm}} 
} 
\caption{Relaxation of average domain length and energy in the ACIC 
model at $T=0.15$ and $T=0.20$.}  
\label{SEde.eps} 
\end{figure}

The Asymmetrically Constrained Ising Chain \cite{jackle}  
is defined in a similar way as the SCIC, 
\numparts
\begin{eqnarray}
E = - \sum_{i=1}^N \sigma_i \\ 
{\cal W} (\sigma_i \rightarrow 1 - \sigma_i) = [1 - \sigma_{i-1}] 
\;\min\{1,e^{-\beta \Delta E}\}
\end{eqnarray}
\endnumparts
The basic difference is the type of constraint used. In this case, 
the class of mobile spins is identified as those for which the left 
neighbour is in the down state. This makes the model more constrained 
than the SCIC, slowing the relaxation dynamics down even more.  
In particular, while in the SCIC a given defect can be reached by an
auxiliary defect from either the left or the right domain, in the 
ACIC this can be done only from the left, due to the asymmetric nature 
of the constraint. This is well illustrated by the relaxation of both 
energy and average domain length, as shown in Fig.\ \ref{SEde.eps}.  
The plateaus present during the relaxation process are a mark of the 
asymmetry of the constraint. They are not present in the SCIC model 
because in that case the system has the freedom to choose the fastest 
way of coalescing domains, meaning that the auxiliary defect will 
naturally travel through the shortest of the two domains adjacent 
to the defect to be eliminated. This produces in the SCIC a slow but  
continuous relaxation of both energy and average domain 
length. In contrast, the ACIC model does not possess the same freedom, 
and domains can grow only leftwards, no matter if this is the fastest 
way of achieving coalescence or not. As a result both energy and average 
domain length display characteristic plateaus corresponding to the time 
needed 
for the flipping of up spins into down spins, related to the
time it takes an auxiliary defect to travel across larger and larger
domains. During this time the system is almost frozen. Of course this
effect is more and more 
noticeable the lower the temperature and results in the typical  
staircase shape for $T=0$ \cite{sollich}. 
 
The different nature of the constraint is also apparent in the time 
scales of the system. In this case only one time scale is present. 
It has been evaluated as $\tau \sim \exp(\beta^2/\lambda)$ with  
$\lambda = \log 2$ in \cite{sollich,mauch}, and has been shown  
to correspond to both correlation and equilibration time \cite{sollich}.    
Note the inverse square temperature dependence in the activated barrier,  
in contrast to the SCIC Model. 

\subsection{The Backgammon (BG) Model} 
 
Let us finally address the last model analysed in this paper, that 
is the Backgammon (BG) Model. The model has been introduced by one of us 
\cite {felix} in 1995. It is defined as 
\numparts
\begin{eqnarray}
E = - \sum_{i=1}^N \delta_{{n_i},0} \\ 
{\cal W} = \min\{1,e^{-\beta \Delta E}\}
\end{eqnarray}
\endnumparts
where $n_i=0,1,...,N$ is the occupation number of each site of  
a $D$-dimensional lattice of $N=L^D$ sites (in the following $D=1$).  
The model can be pictured  
as an ensemble of $N$ particles occupying $N$ boxes, with the energy of the 
system given by the number of empty boxes. Even though this is not strictly 
speaking a kinetically constrained model, an effective constraint is still
present in the form of the conservation of the total number of particles. 
 
Particles can move from one box to 
another one with a probability given in terms of temperature 
by the ordinary Metropolis factor. Once the starting box is specified,  
different choices can be made on how to select 
the arrival box. In the original paper \cite{felix} both the starting and  
the arrival box were chosen randomly. The relaxation of the model, 
characterized by a mean field dynamics which turned out to be exactly  
solvable \cite{bg-closed}, was then proven to rest on the overcoming of  
entropic barriers. 
In contrast we assume here a dynamics  
where particles can move only to nearest-neighbour boxes, introducing  
thereby activated processes as relevant processes  
in the relaxation properties of the system.  
This change is expected to introduce a coarsening behaviour, which was 
absent in the original model.  
 
\begin{figure}[h] 
\hbox{ 
\centerline{\epsfig{file=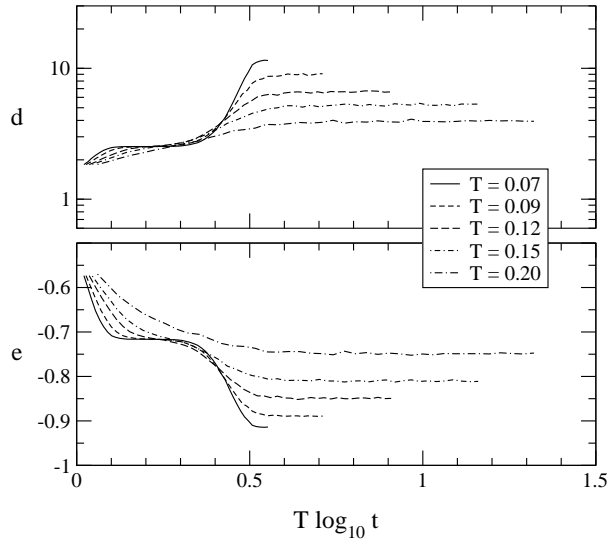,width=8cm}} 
}  
\caption{Average domain length and energy in the BG Model. The plateau  
appearing at lower temperatures is related to the microscopic time 
scale $\tau_1$ and is representative of the time spent by the system  
in the entropic elimination of defects.}
\label{BGde}   
\end{figure}

More specifically, after an initial fast evolution of the system, a   
situation will be achieved where multiply occupied boxes are separated 
by empty boxes and singly occupied boxes ({\em defects}). Then two 
types of microscopic processes can take place. A first possibility is the 
wandering of a defect till it gets to a multiply occupied box. During 
this process no change in the energy of the system will occur, 
implying that the process is entropically driven. In fact, 
the decrease of the energy when the defect sticks to a multiply 
occupied box is related to the discovery of the right path in the 
configuration space to get to that multiply occupied box. This process 
is clearly entropic. On the other hand the annihilation of two 
multiply occupied boxes also requires activation. In this case the 
creation of a defect is involved, and this is an activated process 
since an empty box must be occupied, leading thereby to an increase of 
the energy. As a result, this version of the model shows a coarsening 
behaviour related to the increase of the size of {\em domains of empty boxes}. 
In Fig.\ \ref{BGde} we report the  
behaviour of energy and average domain length at different temperatures.  
It is easy to show that also for this model $d$ and $e$ are related via
$d = -e/(1+e)$.
 
According to the presence of the two processes mentioned above,  
two time scales are present in the system. The first time scale is  
associated with the entropic mechanism and is estimated in \cite{crrs}  
as $\tau_1 \sim \exp(\beta)/\beta$. This time scale plays a similar role as 
the time scale $\tau_1$ defined in the SCIC Model. The equilibration of the 
system, on the other hand, proceeds via the activated mechanism
described above  
and is estimated again in \cite{crrs} as $\tau_{\rm eq} \sim \beta \exp 
(\beta)$. The interplay between these two different time scales is 
shown in Fig.\ \ref{BGde}.  
 
Finally let us remark that this model, too, clearly shows glassy 
dynamics since after the initial elimination of defects the 
successive elimination of multiply occupied boxes becomes slower and slower  
as time goes on, and this effect increases exponentially as the temperature  
is lowered.

\section{Fluctuation-Dissipation Relation} 
 
In the previous section we have seen that the models under study 
are characterized by different dynamics. All of them are 
constructed in such a way as to exhibit both coarsening and glassy 
behaviour. Nevertheless the different nature of the constraints 
inserted results in the existence of different time scales,  
and produces different relaxation features.    
 
In order to get more insights into the different dynamics of these 
models we analyse their response to an external perturbation.  
An efficient way of doing this is through the so called fluctuation  
dissipation plots \cite{cuglia}, where the response is plotted as a function 
of the correlation.    
 
First of all, we need to define a suitable perturbation. This must be 
chosen in such a way that the linear response regime applies and 
also it must be not coupled with the absorbing state, that is the 
ground state. A good choice is the following: 
\be 
\delta {\cal H}(t) = - h_0 \Theta(t - t_w) \sum_{i=1}^N \epsilon_i 
\sigma_i. 
\ee 
Here $h_0$ is a (small) constant external field, and $\epsilon_i$ are just 
zero mean random quenched variables which can take the values 
$\pm1$. After an initial free evolution starting from a random 
configuration, the perturbation is turned on at time $t_w$. The spin 
variables are the usual ones defined in the SCIC and ACIC models and 
are defined as $\sigma_i=\delta_{n_i,0}$ in the BG model.   
 
Accordingly we measure the correlation function, 
\be 
C(t+t_w,t_w)=\frac{1}{N}\sum_{i=1}^N\nu_i(t_w)\,\nu_i(t+t_w), 
\ee 
and the staggered magnetization, 
\be 
M_{\rm stag}(t+t_w,t_w)=\frac{1}{N}\sum_{i=1}^N\epsilon_i\,\nu_i(t+t_w). 
\ee 
For reasons that will become clear shortly we used the variables $\nu_i = 
2 \sigma_i - 1$ in place of the $\sigma_i$'s. 
 
In general, at equilibrium, for any two times $t$ and $t^{\prime}$ 
correlation $C$ and response $R$ are related by the Fluctuation 
Dissipation Theorem (FDT) as 
\be 
R(t - t^{\prime}) = \beta \frac{\partial C(t - t^{\prime})}{\partial 
t^{\prime}}, \label{fdt1} 
\ee 
where the explicit dependence on the two times is lost due to the 
invariance under time translation at equilibrium. Defining the
integrated response function as 
\be 
\chi(t - t^{\prime}) = \int_{t^{\prime}}^t du\ R(t,u), 
\ee 
Eq.\ (\ref{fdt1}) can be rewritten as  
\be 
\chi(t - t^{\prime}) = \beta[C(0) - C(t - t^{\prime})]  
= \beta [1 - C(t - t^{\prime})], \label{fdtint} 
\ee 
where the second equality holds when the variables $\nu_i$'s are 
assumed. Then plotting $T\chi$ as a function of $C$ will result in a 
straight line with slope $-1$. 
 
Of course these properties are not expected to be valid if the 
regime under investigation is out of equilibrium. First of all we expect 
that correlations and responses will be generally dependent   
on the two separate times $t$ and $t^{\prime}$. Secondly, explicit 
violations of FDT will have to show up in Eq.\ (\ref{fdt1}). A 
parametrization of such violations has been proposed in 
\cite{cuglia}, and consists in generalizing Eq.\ (\ref{fdt1}) to
\be 
R(t,t^{\prime}) = \beta_{\rm eff} \frac{\partial  
C(t,t^{\prime})}{\partial t^{\prime}} \label{fdt2}, 
\ee 
where $\beta_{\rm eff} = \beta(C) = \beta X(C)$ is interpreted as an effective 
temperature. The corresponding integral representation of (\ref{fdt2}) 
is   
\be 
\chi(t,t^{\prime}) = \int_{C(t,t^{\prime})}^1 \beta(C) \ dC = \beta  
\int_{C(t,t^{\prime})}^1 X(C) \ dC. 
\ee  
For equilibrium dynamics, $X(C) = 1$ and Eq.\ (\ref{fdtint}) is recovered, 
while violations will appear for offequilibrium behaviour, manifesting  
themselves as deviations from the straight line with slope $-1$ of the 
corresponding equilibrium regime.  
  
\begin{figure}[h]   
\hspace*{-1.2cm}
\begin{minipage}{8cm}   
\begin{center}
\epsfig{file=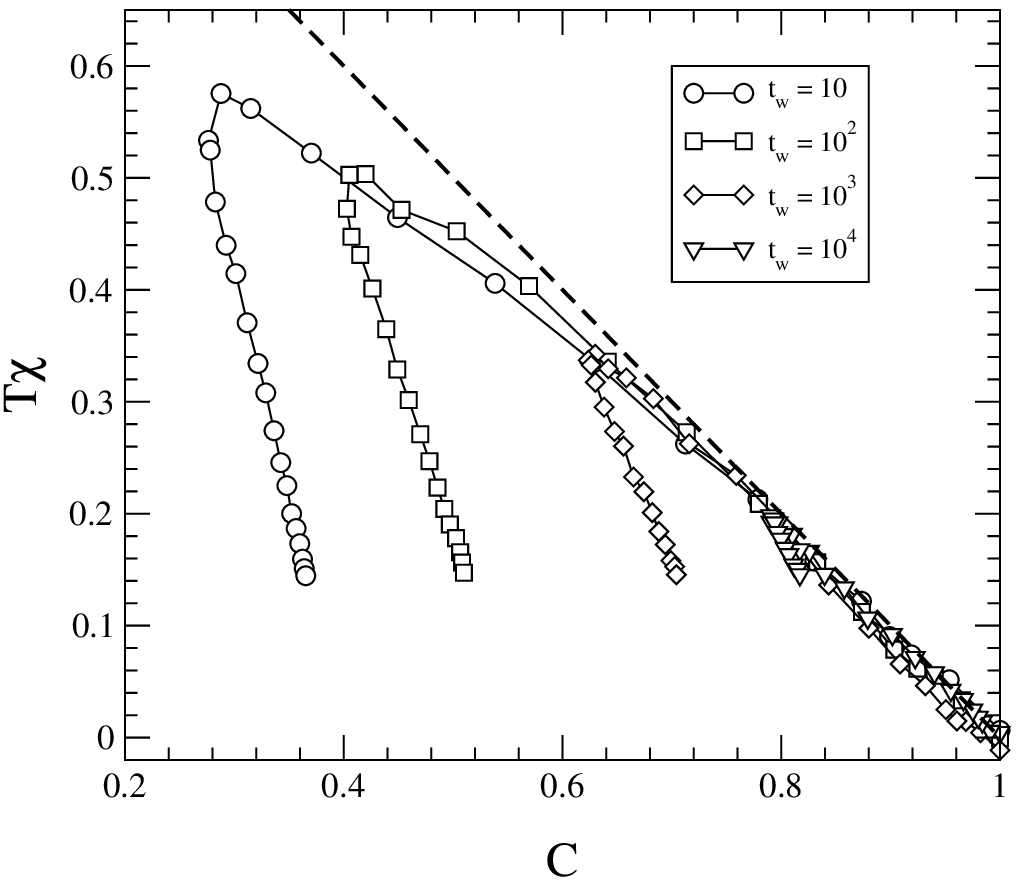,width=6.5cm} 
\caption{FDT plots in the SCIC for $N=10^5$, $T=0.3$ and different  
waiting times $t_w=10,100,1000,10000$. The straight line is the FDT     
relation.} 
\label{FAfdt.0.30} 
\end{center}
\end{minipage}  
\hspace{-1.5cm}
\begin{minipage}{8cm}  
\begin{center}
\epsfig{file=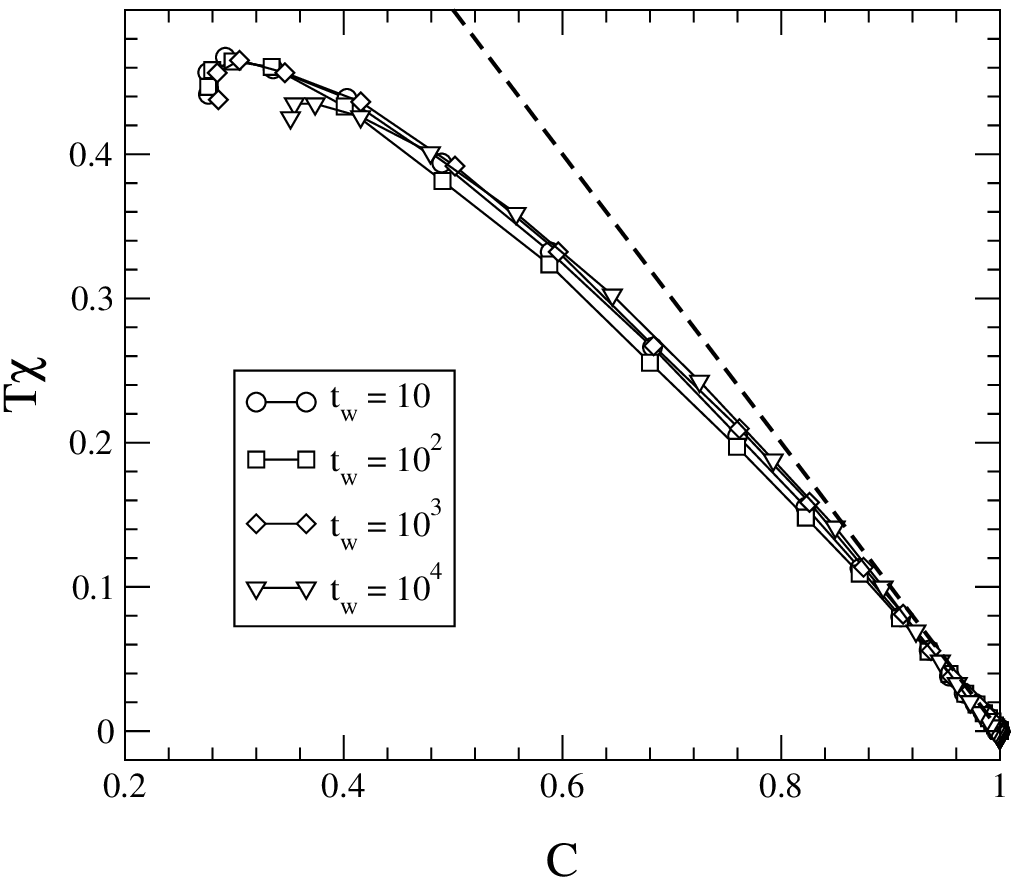,width=6.5cm} 
\caption{FDT plots in the SCIC for $N=10^5$, $T=0.11$ and different 
waiting times $t_w=10,100,1000,10000$. The straight line is the FDT 
relation.} 
\label{FAfdt.0.11} 
\end{center}
\end{minipage}  
\end{figure} 
We show in Fig.\ \ref{FAfdt.0.30} and \ref{FAfdt.0.11} two plots            
of the integrated response function $T \chi = T M_{\rm stag}/2 h_0$ 
as a function of the correlation for the SCIC model and for  
two different temperatures.                                                 
The existence of different activated relaxation times results in  
rather peculiar FDT plots. For $t_w<\tau_1$ (Fig.\ \ref{FAfdt.0.11})  
$C$, $\chi$ and $X$ do not show any dependence on $t_w$,  
nevertheless $X$ is a non-trivial function of $C$ corresponding to  
nonequilibrium behaviour without aging. For $t_w > \tau_1$, Fig.\  
\ref{FAfdt.0.30}, there are aging effects and $X$ shows the typical  
two slope pattern. However, the existence of a second typical time 
scale results in a second downwards bending of the integrated response 
function and $X$ as function of $C$ has a three slope shape. 
 
\begin{figure}[h] 
\hspace*{-1.2cm}
\begin{minipage}{8cm} 
\begin{center}
\epsfig{file=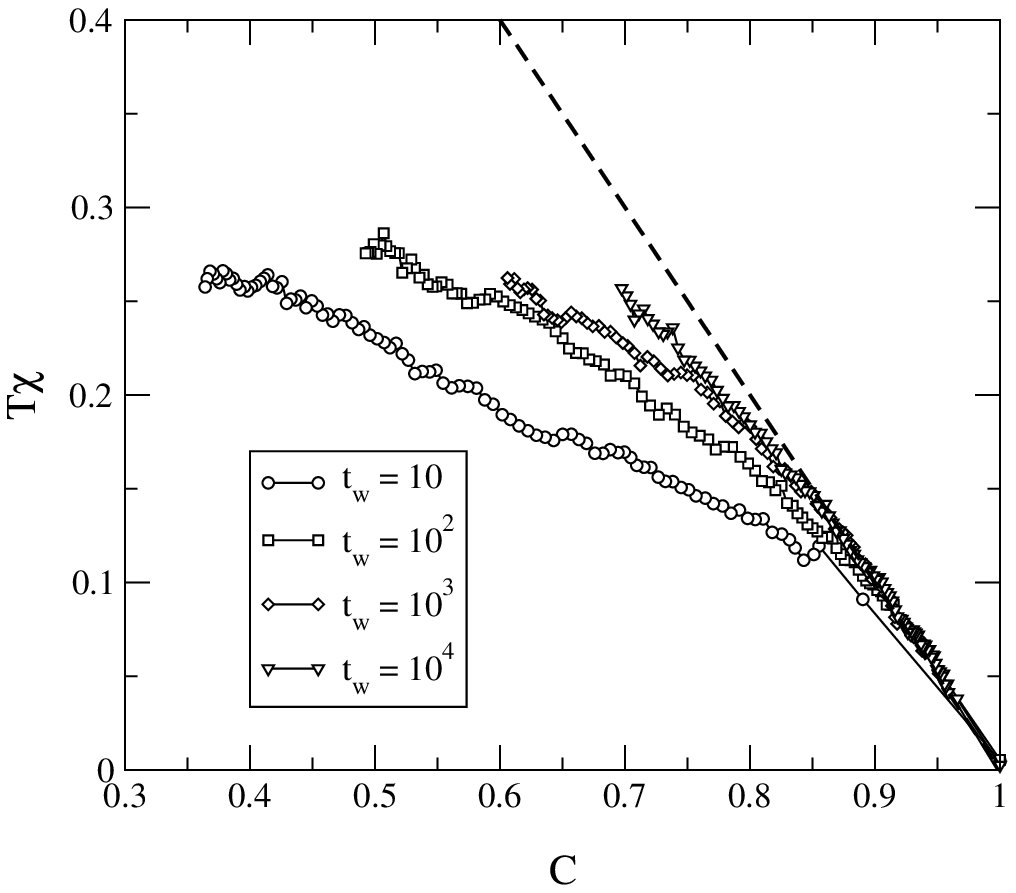,width=6.5cm}
\caption{FDT plots in the ACIC for $N=10^5$, $T=0.4$ and different 
waiting times $t_w=10,100,1000,10000$. The straight line is the FDT 
relation.} 
\label{SEfdt.0.40} 
\end{center}
\end{minipage} 
\hspace{-1.5cm}
\begin{minipage}{8cm} 
\begin{center}
\epsfig{file=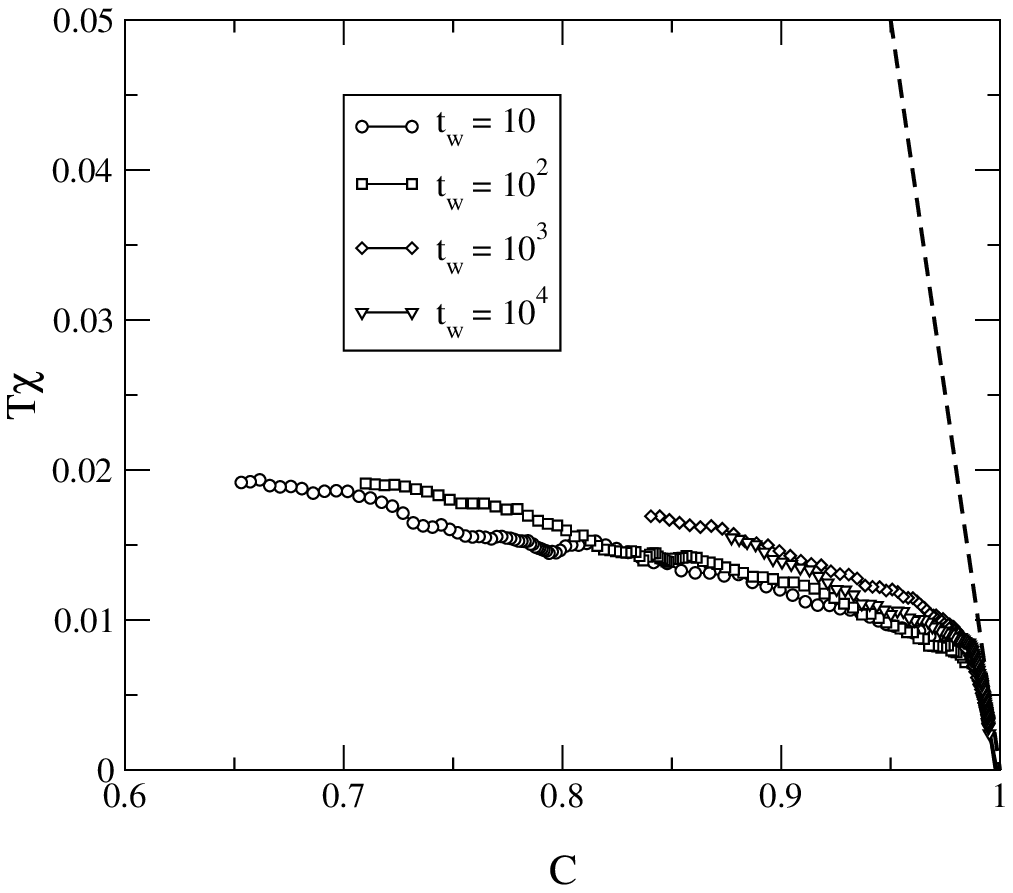,width=6.5cm}
\caption{FDT plots in the ACIC for $N=10^5$, $T=0.2$ and different 
waiting times $t_w=10,100,1000,10000$. The straight line is the FDT 
relation.} 
\label{SEfdt.0.20} 
\end{center}
\end{minipage} 
\end{figure}

We repeat the same analysis for the ACIC Model.  
In Fig.\ \ref{SEfdt.0.40} and \ref{SEfdt.0.20} we show the FDT plots 
for the ACIC at temperatures $T=0.4$ and $T=0.2$ respectively. 
Interestingly, for waiting times comparable with the correlation time,  
so that the system is not too far from equilibrium, the 
fluctuation-dissipation ratio $X$ rapidly converges to $1$, 
see Fig.\ \ref{SEfdt.0.40}. At low temperatures (Fig.\ \ref{SEfdt.0.20}),   
$t_w\ll \tau_{\rm corr}$ and the fluctuation-dissipation ratio is very 
small, $X\simeq 0.1$, and roughly independent of $t_w$, a scenario 
typical of coarsening models \cite{barrat}. 
 
\begin{figure}[h]
\hspace*{-1.2cm}
\begin{minipage}{8cm} 
\begin{center}
\epsfig{file=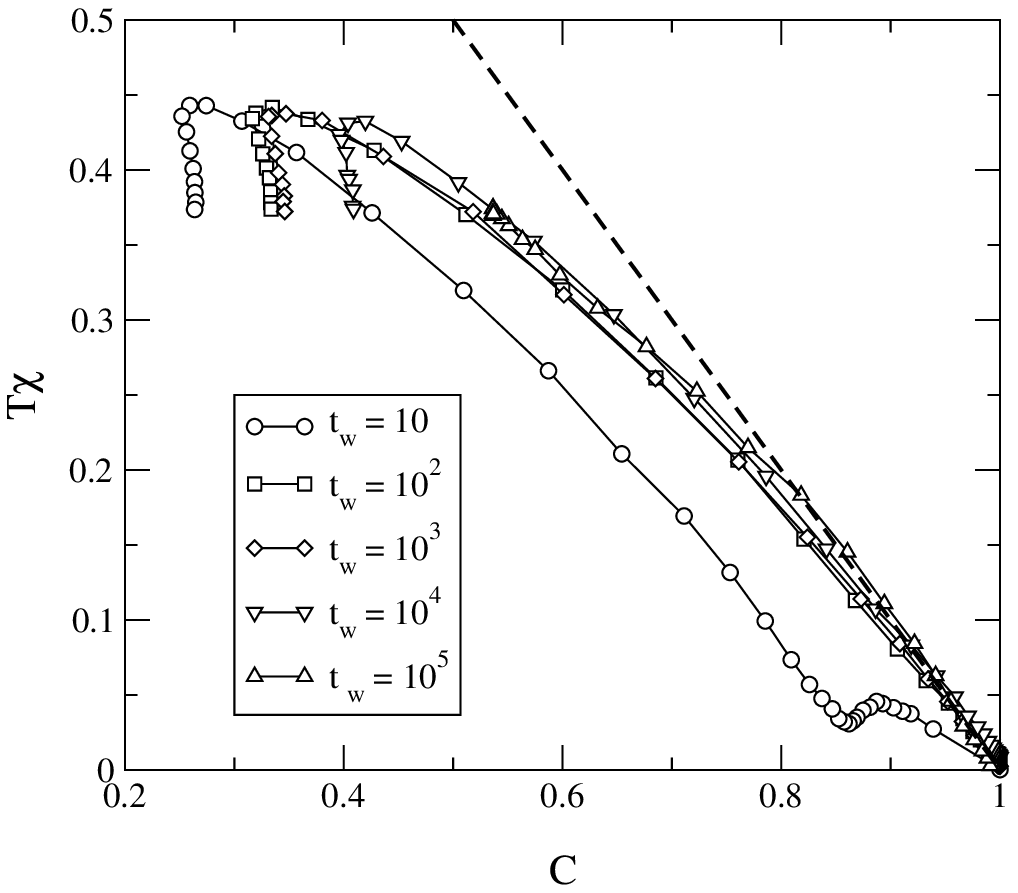,width=6.5cm} 
\caption{FDT plots in the BG for $N=10^4$, $T=0.1$ and different values 
of $t_w$.} 
\label{BGfdt.0.10} 
\end{center}
\end{minipage} 
\hspace{-1.5cm}
\begin{minipage}{8cm} 
\begin{center}
\epsfig{file=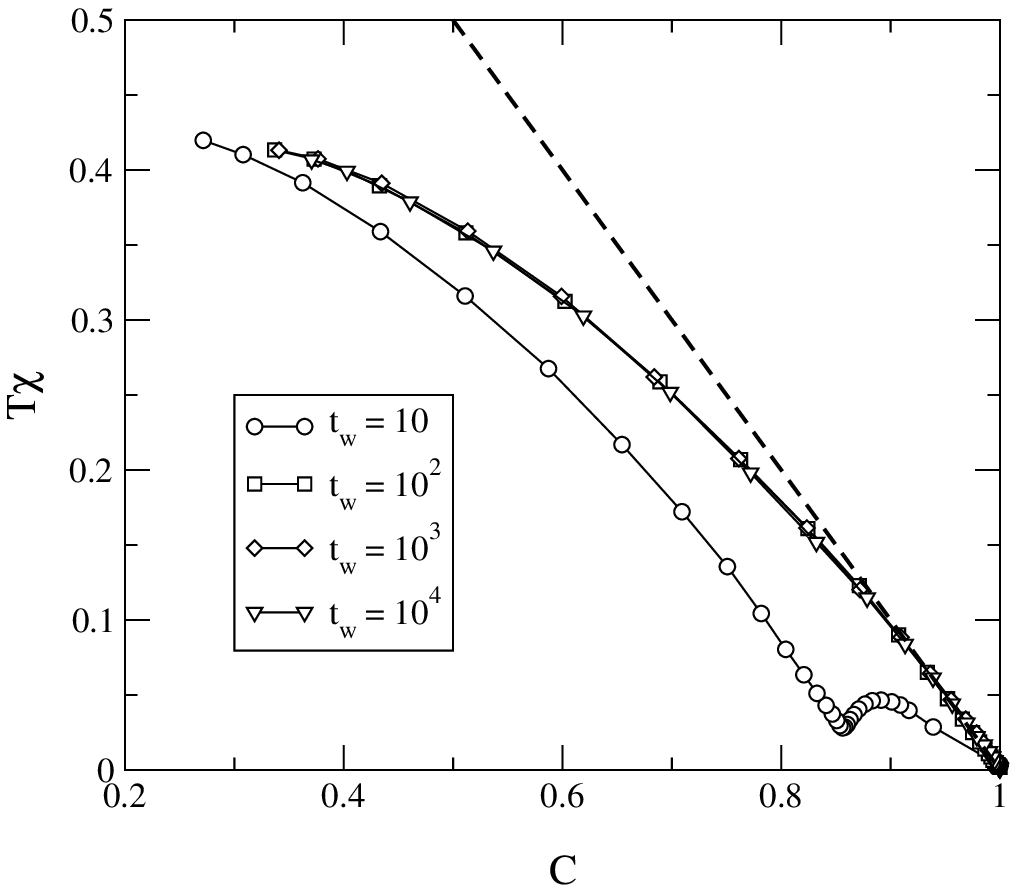,width=6.5cm} 
\caption{The same as Fig.\ \protect\ref{BGfdt.0.10} for $T=0.09$.} 
\label{BGfdt.0.09} 
\end{center}
\end{minipage} 
\end{figure}

Finally we address the BG Model. Our results are shown in Fig.\ 
\ref{BGfdt.0.10} and \ref{BGfdt.0.09}.  
Note that aging effects are absent for $t_w < \tau_1$, but
nevertheless $X < 1$.
For waiting times $\tau_1<t_w< \tau_{\rm eq}$ the system 
shows strong non-equilibrium effects with a downwards bending of the 
integrated response function as a function $C$, similar to 
what is seen in the SCIC model. The origin of this effect is, however,
different 
and follows from the asymmetric response of  
occupied and empty boxes to the staggered field. Since the field is
coupled to empty boxes,
the typical time to empty a box is larger than that  
to occupy an empty one. In other words, when quenching from high  
(or infinite) temperature, boxes are occupied fast and their number converges  
relatively fast towards the equilibrium value. However, due to the staggered  
field, the distances between them are far from the equilibrium value and  
occupied boxes must be rearranged, which is a very slow process. 
 
\vspace{0.5cm}

\section{The Stillinger and Weber decomposition} 
 
\begin{figure}[h] 
\hbox{ 
\centerline{\epsfig{file=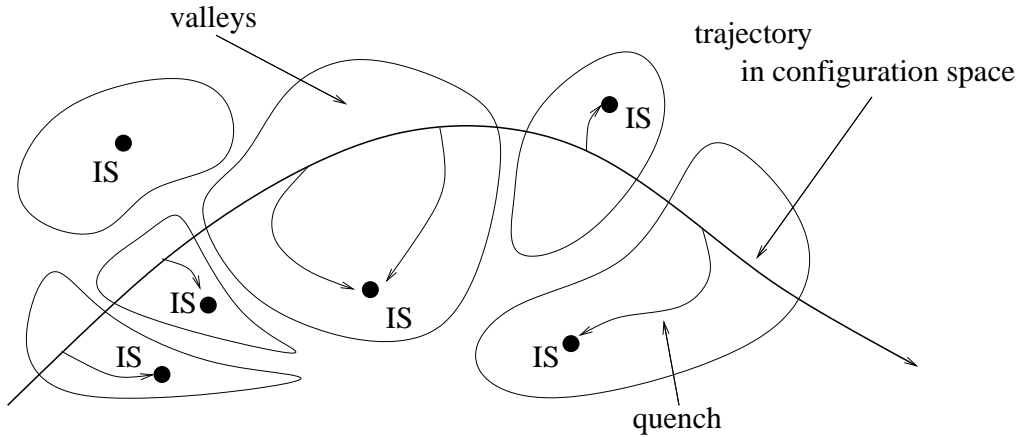,width=14cm}} 
} 
\caption{A pictorial description of the Stillinger and Weber 
decomposition. Equilibrium configurations are regularly quenched to 
reach the corresponding minimum of the phase space (inherent structure).  
The set of configurations reaching the given minimum is called 
basin of attraction or valley of that minimum.} 
\label{sw} 
\end{figure}

An interesting approach to the investigation of activated behaviour in
glasses
was suggested in the Eighties by Stillinger and Weber \cite{sw}.  
As shown in Fig.\ \ref{sw}, each configuration of the 
system is mapped into a local minimum of the energy through a local 
potential energy minimization ({\em quench}) starting from the  
given configuration. The local minimum was called {\em Inherent 
Structure} (IS), while the set of configurations flowing into it 
defines the {\em basin of attraction} or {\em valley} of the IS. 

Following SW one constructs an IS-based thermodynamics decomposing  
the partition function into a sum over IS with the same energy  
\cite{sw} 
\be 
{\cal Z}(T) \simeq \sum_{e} {\cal Z}_{IS}(e,T), \label{partfunc} 
\ee 
with  
\be 
{\cal Z}_{IS}(e,T) = \exp [N (-\beta e + s_c(e) -\beta 
f(\beta,e))]. \label{eqpe}  
\ee 
Here $s_c(e)$ is the configurational 
entropy, which yields the number of different IS with energy $e$: 
\be 
\Omega(e)=\exp(Ns_c(e)). 
\ee 
The term $f(\beta,e)$ accounts for the free energy of 
the IS-basin of energy $e$, {\em i.e.},  
the partition sum restricted to the basin of attraction of IS with energy  
$e$. In each IS-basin the energy has been shifted, so that the IS  
has zero energy, and $f$ accounts only for energy differences. Then 
the probability of finding an IS with energy $e$ is given by the 
expression: 
\be 
{\cal P}_{IS}(e,T)=\exp [N (-\beta e + s_c(e) -\beta f(\beta,e))]/{\cal Z}(T). 
\ee 
 
In general $f(\beta,e)$ may have a non-trivial dependence on the 
energy if the IS-basin of IS with different energy are different.  
Usually it is reasonable to expect that $f(\beta,e)$ is roughly 
independent of $e$ at least in two different situations. The first is 
when the temperature is such that only the states near the bottom of the 
IS-basin contribute \cite{skt,crisanti}, and the second is when the  
IS-basins are narrow and contain few configurations, as in  
REM-like models \cite{crisanti,derrida}.   
 
When the $e$-dependence of $f$ can be neglected, the configurational 
entropy $s_c(e)$ can be obtained directly from (\ref{eqpe}). From an 
operative point of view, the procedure that we followed consists in 
the following steps:  
\begin{enumerate} 
 
\item We equilibrate the system at temperature $T$ with  
$t_{\rm therm}$ Monte Carlo steps (MCS). 
 
\item We run a group of $t_{\rm run}$ MCS. At the end of the group  
we perform a steepest descent procedure ($T=0$ Monte Carlo dynamics)  
to identify an IS.  
 
\item We repeat step 2 $N_{\rm run}$ times. 
 
\item We keep in memory the number of times $N_{\rm IS}$  
we have found an IS with a given energy $e$. 
 
\item We construct the histogram ${\cal P}_{IS}(e,T)$. 
 
\item We calculate $s_c(f,T)$ as 
\be 
s_c(e) = \beta e + \frac{1}{N} \log {\cal P}_{IS}(e,T) + \Delta(T), 
\label{fsc} 
\ee 
where  
\be 
\Delta(T) = \beta f(\beta) + \frac{1}{N} \log {\cal Z}(T) 
\ee  
is assumed to be a function of temperature only, and is computed by
imposing the collapse of the data points onto a single curve. 
 
\end{enumerate} 
 
The substitution of the original partition function (\ref{partfunc})  
with the sum of the partion functions in each valley is expected to be valid  
and to reproduce the correct 
thermodynamics since it corresponds simply to a different way of 
summing the partition function. Of course this is true within the 
approximation that all the valleys have the same relevance to the  
statistical properties of the system. The scenario that we are 
proposing rests on the idea that to describe the equilibrium 
properties of the system it is sufficient to count its IS.  
In other words we implicitly assume the existence of an equiprobability  
principle working for the IS themselves, according to which the  
frequency of visit of an IS with a given energy is only dependent  
on the total number of IS present in the system at that energy. Then the 
configurational entropy can be considered as analogous to the ordinary 
Boltzman entropy, which in the Gibbs ensemble counts the number of  
configurations with a given energy. In contrast, if the frequency of 
visit of a given IS is also dependent on different parameters, then such  
a construction may not work. An example of this is the 
Sherrington-Kirkpatrick Model, where the size of the basins must also 
be taken into account. In this case, for the configurational entropy to be 
meaningful it needs to be expressed in terms of the free energy of the
valleys, not in terms of their potential energies \cite{marinari}.
 
However, the definition of the SW projection is also dynamical in its 
own nature, and contains relevant information about the offequilibrium 
properties of the system. This is due to the intrinsically dynamical way of 
defining the IS themselves, which is through a quenching procedure 
based on the dynamics of the system. This partially answers the criticism  
recently raised by Monasson and Biroli \cite{biroli} about the definition  
itself of the IS. On the other hand, that criticism remains meaningful in 
that it highlights how the configurational entropy, even though well defined,  
may not be able to capture the relevant dynamics of the system. According to  
its definition, the configurational entropy not only contains information  
about the equilibrium properties of the system, but also on how the system  
approaches equilibrium. As we shall see, this is the key point that makes 
the SW decomposition unsuitable for coarsening systems such as the  
ones discussed in this paper. 
 
We have calculated the configurational entropy analytically for all 
the models presented. For both the SCIC and the ACIC it is possible to show 
\cite{crrs} that the zero temperature dynamics can be solved exactly 
and that ${\cal P}_{IS}(e,T)$ has the form
\be 
{\cal P}_{IS}(e,T)=\frac{1}{\sqrt{2\pi\langle C_0^2(\infty)\rangle_c}}\exp
\Bigl ( -\frac{(e-\langle e_{IS}\rangle)^2}{2\langle 
C_0^2(\infty)\rangle_c}\Bigr), \label{pis} 
\ee 
where $\langle e_{IS} \rangle$ and $\langle C_0^2(\infty) \rangle_c$ are 
known in terms of the equilibrium magnetization (see \cite{crrs}).  
Also counting the number of fixed points of the dynamics can produce
an estimate of the configurational entropy, which results in \cite{crrs}
\be 
s_c(e) &=& \frac{\log(N_{fix})}{N} \nonumber \\
&=&-e\log(-e) -(1+e)\log(1+e) +(1+2e)\log(-1-2e). 
\label{eqs2} 
\ee 
The main point to highlight here is that both these results, Eq.\ (\ref{pis}) 
and (\ref{eqs2}), are the same for the SCIC as well as the ACIC 
Model. As a consequence we expect the two models to have the same  
configurational entropy.  
 
In the case of the BG Model, the zero temperature dynamics cannot be 
closed exactly. However we can still carry out an estimate of the 
configurational entropy by counting the number of fix-points. In this 
case we obtain \cite{crrs}: 
\be 
s_c(e) = &-&(1+e)\log(1+e)-e\log(-e)+(1+2e)\log(-1-2e) \nonumber \\
&-& \log(y)+(1+e) \log(\exp(y)-y-1) \label{eqs4}  
\ee 
where $y$ satisfies the saddle-point condition, 
\be 
e=-1+\frac{\exp(y)-1-y}{y(\exp(y)-1)}. 
\ee 
Note that for the BG Model the configurational entropy may be 
negative because particles are distinguishable.  
 
\begin{figure}[h] 
\hspace*{-1.2cm}
\begin{minipage}{8cm} 
\begin{center}
\epsfig{file=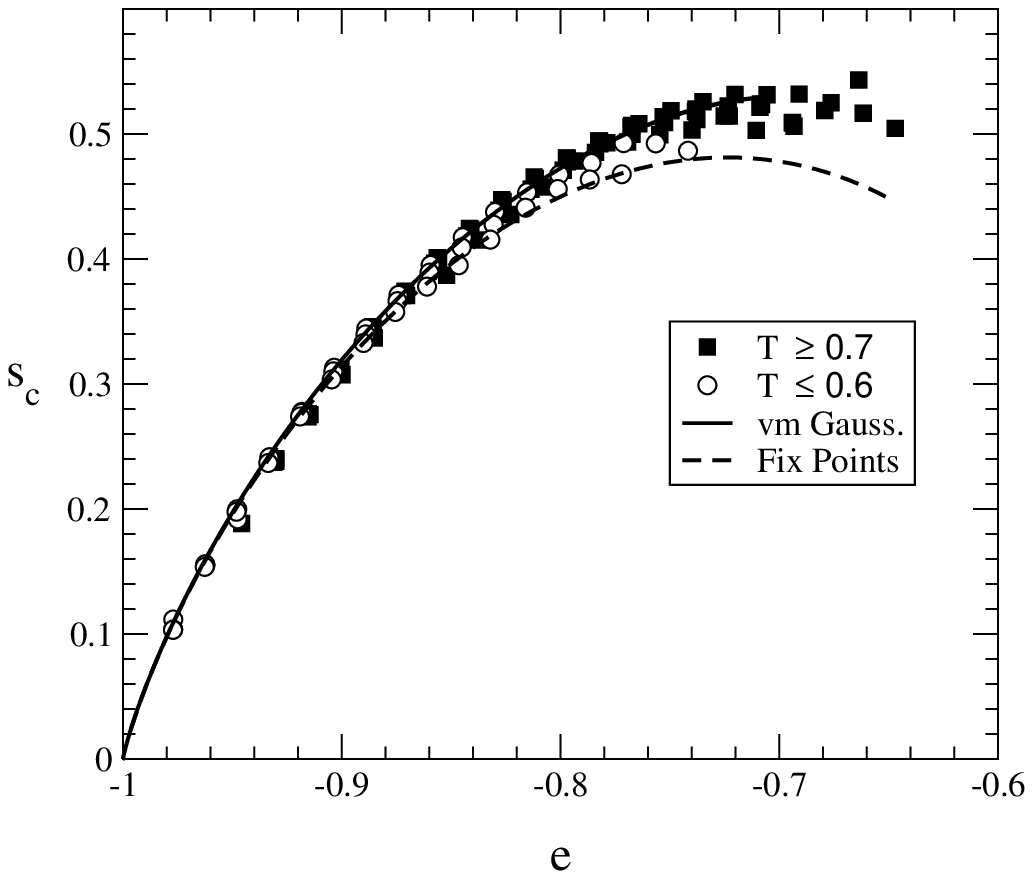,width=6.5cm} 
\caption{SW configurational entropy in the SCIC for $N=64$ spins at 
different temperatures compared with the analytical prediction 
(\ref{sc1}) (upper curve) and the fix-point estimate (\ref{eqs2}) 
(lower curve).} 
\label{FAsc} 
\end{center}
\end{minipage} 
\hspace{-1.5cm}
\begin{minipage}{8cm} 
\begin{center}
\epsfig{file=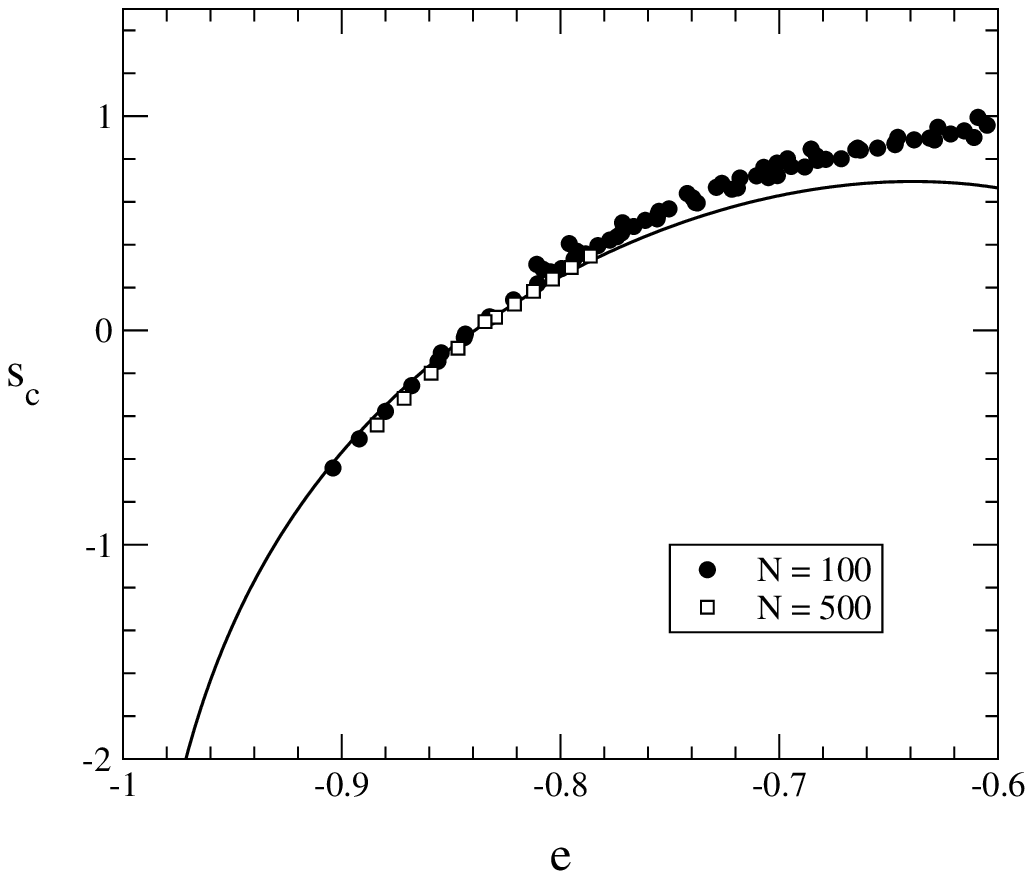,width=6.5cm}
\caption{SW configurational entropy in the BG model for $N=100,500$ 
boxes at different temperatures $T=1.0,0.5,0.4,0.3,0.2,0.15,0.1$  
compared with the fix-point estimate (\ref{eqs4}) (full line).} 
\label{BGsc} 
\end{center}
\end{minipage} 
\end{figure}

We show both the analytical predictions and the numerical data in
Figs. \ref{FAsc} and \ref{BGsc}. In Fig.\ \ref{FAsc} we show the
results obtained for the SCIC 
model with $N=64$ and different temperatures. The SW configurational entropy  
$s_c$ is obtained from the numerical ${\cal P}_{IS}(T,e)$ as in 
eq.\ (\ref{fsc}). For each temperature $\Delta(T)$ has been fixed by   
collapsing different data onto a single curve. 
As a comparison we also show the theoretical 
predictions from equations (\ref{eqs2}) and (see \cite{crrs}) 
\be 
s_c(e)=\int_{0}^T\frac{d\langle e_{IS}\rangle}{dT}\frac{dT}{T}. 
\label{sc1} 
\ee 
As shown in \cite{crrs}, both coincide asymptotically 
close to the ground state energy $e=-1$. The collapse is excellent,
showing that the approximation (\ref{eqs2}) and the low-temperature 
behaviour (\ref{sc1}) asymptotically coincide in the limit $T\to 0$.  
We note that there is a range of energies where data from  
$T\le 0.6$ collapse onto one curve while data for higher temperature  
collapse onto a different curve. This residual temperature dependence  
follows from the presence of many equivalent directions for  
energy minimization \cite{crrs}. We have checked that ${\cal P}_{IS}(e,T)$  
is the same for the ACIC Model. In all cases we find the same  
results. 
 
In Fig.\ \ref{BGsc} we show the numerical computation of $s_c$ for the 
BG Model, for two different sizes $N=100,500$ and temperatures ranging  
from $T=0.1$ up to $T=1$. Similarly to what we found for 
the constrained kinetic models, the data collapse nicely onto a 
single curve although they do not exactly coincide with prediction
derived from the number of fixed 
points. In this model the presence of different equivalent directions for
decreasing the energy does not influence $s_c$. This is most probably
due to the global character of the constraint. 
 
Comparing Figs. \ref{FAsc} and \ref{BGsc} we see that 
the agreement between measured and predicted configurational entropy 
is now worse. We attribute this to the 
presence of entropic barriers which follows from all 
possible arrangements of particles inside the 
boxes. All arrangements leave the energy unchanged, but their 
number strongly depends on the number of empty boxes, leading 
to a stronger energy dependence of the IS free energy for this model. 
This effect is not present in the kinetically constrained Ising chain. 
 
The conclusion that can be drawn from this section is that for the
models considered a description of their glassy behaviour in terms of
a complex energy landscape is not relevant. Even though the SW
configurational entropy for the constrained Ising chain is a
non-trivial quantity, it does not distinguish the SCIC model from the
ACIC model.

\section{Concluding Remarks} 
 
In this paper we have analysed the possibility of decomposing the 
dynamics of $1D$ constrained models according to the 
Stillinger and Weber prescription \cite{sw}.  
 
In particular we have focussed on the Symmetrically and Asymmetrically  
Constrained Ising Chains \cite{fred,jackle} and on the Backgammon 
Model \cite{felix}. All these Models have been proven to exhibit 
glassy and coarsening behaviour. Their approach to equilibrium has 
turned out to be quite different, due to 
the different microscopic kinetic constraints. This was 
apparent specifically in their different Fluctuation Dissipation Plots. 
 
In contrast the SW projection always results in
configurational entropies with the same qualitative features. 
We argue that this is related to the 
fact that a growing length scale is present in these systems, driving 
the equilibration processes. In other words, when one substitutes the 
original dynamics with an IS-based dynamics, one is not able to 
transfer to the IS level the information about the correlation between the 
successive configurations reached during the 
approach to equilibrium. If the system under consideration evolved 
between uncorrelated configurations, then the SW approach would be 
powerful, as has been proved in other cases 
\cite{crisanti,marinari}. However, here the missing 
coarsening-biased choice in the jumps between an IS and the following 
one prevents it from working properly.  
 
The SW approach is expected to hold for systems where the relevant  
equilibration is driven by an entropic process, with activated jumps  
between different basins occurring with the same probability \cite{felix2}. 
 
Discerning a good class of simple and tractable models which contain 
the relevant mechanisms responsible for relaxation in real glasses  
would be a very important step in the direction of building a microscopic 
theory for the glass transition beyond ideal mode-coupling theory.


\begin{thebibliography}{99}
\bibitem{nota} Note that the definition here adopted for $d$ is 
different from the one adopted in \cite{sollich}. In our case, we do 
not consider the border of the domains in the computation of their 
size. The two definitions differ by one unit.
\end{thebibliography}
\end{document}